%% file: ms_revise3.tex
\documentclass[manuscript]{aastex}

\begin{document}
\title{Coronal Magnetic Fields Derived from Simultaneous Microwave and EUV Observations and Comparison with the Potential Field Model}
\author{Shun Miyawaki}
\affil{Department of Science, Ibaraki University, Mito, Ibaraki 310-8512, Japan}
\email{shunmi089@gmail.com}
\author{Kazumasa iwai\altaffilmark{1} and Kiyoto Shibasaki}
\affil{Nobeyama Solar Radio Observatory, National Astronomical Observatory of Japan, Minamimaki, Nagano 384-1305, Japan}
\author{Daikou Shiota}
\affil{Solar-Terrestrial Environment Laboratory, Nagoya University, Nagoya, Aichi 464-8601, Japan}
\and
\author{Satoshi Nozawa}
\affil{Department of Science, Ibaraki University, Mito, Ibaraki 310-8512, Japan}

\altaffiltext{1}{National Institute of Information and Communications Technology, Koganei 184-8795, Tokyo, Japan}
\begin{abstract}
We estimated the accuracy of coronal magnetic fields derived from radio observations by comparing them to potential field calculations and the DEM measurements using EUV observations. We derived line of sight component of the coronal magnetic field from polarization observations of the thermal bremsstrahlung in the NOAA active region 11150, observed around 3:00 UT on February 3, 2011 using the Nobeyama Radioheliograph at 17 GHz. Because the thermal bremsstrahlung intensity at 17 GHz includes both chromospheric and coronal components, we extracted only the coronal component by measuring the coronal emission measure in EUV observations. In addition, we derived only the radio polarization component of the corona by selecting the region of coronal loops and weak magnetic field strength in the chromosphere along the line of sight. The upper limit of the coronal longitudinal magnetic fields were determined as 100--210 G. We also calculated the coronal longitudinal magnetic fields from the potential field extrapolation using the photospheric magnetic field obtained from the Helioseismic and Magnetic Imager (HMI). However, the calculated potential fields were certainly smaller than the observed coronal longitudinal magnetic field. This discrepancy between the potential and the observed magnetic field strengths can be explained consistently by two reasons; (a) the underestimation of the coronal emission measure resulting from the limitation of the temperature range of the EUV observations, (b) the underestimation of the coronal magnetic field resulting from the potential field assumption.

\end{abstract}

\section{INTRODUCTION}
Solar activities in the corona, such as flares, jets, eruptions, and heatings, are dominated by magnetic fields. Therefore, solar phenomena can be most appropriately understood by measuring the coronal magnetic field through phenomena such as polarization generated by the Zeeman effect and/or the Hanle effect in the optical range. However, such effects are obscured by the line broadening of the high temperature plasma, and the low polarization degree of the coronal weak magnetic field. \citet{Lin2004} measured coronal magnetic fields to a few gauss at a spatial resolution of 20$\arcsec$ and 70 minutes of integration, but this method is still developing.

The coronal magnetic field also can be derived by extrapolating the photospheric magnetic field from the potential or force-free field approximations \citep{Sakurai1989,Shiota2008,Shiota2012}. Generally, the magnetic pressure in the corona is thought to exceed the gas pressure, implying that coronal plasma has a low beta. Therefore, the coronal magnetic field should be near to the potential or force-free. Structurally, the coronal loops observed by EUV often resemble the magnetic field lines derived by extrapolations \citep{Wiegelmann2012}. However, the magnetic field in the photosphere is not considered as force-free because of the high beta of photospheric plasma \citep{Gary2001}. In addition, the magnetic field strengths observed in different spectral lines are not identical because of the differences in magnetic sensitivities and formation heights of the spectral lines. Hence, the coronal magnetic field derived by extrapolated from the photospheric magnetic field should be validated with other observations.

Several methods derive the magnetic field from microwave observations. For instance, the strong magnetic field above sunspots can be derived from gyro-resonance emission \citep{White2004,Bogod2012}. Other methods derive the coronal magnetic field from polarization reversal by quasi-transverse propagation \citep{Ryabov2005}. \citet{Brosius1997} estimated the coronal magnetic field by fitting the observed radio intensity to the radio intensity calculated from EUV observations.

The circular polarization degree of thermal bremsstrahlung (also called free-free) radiation depends on the longitudinal magnetic field strength (hereafter, we refer to the longitudinal magnetic field as the magnetic field along the line of sight). \citet{Bogod1980} measured the magnetic field strength in the upper chromosphere as approximately 40 G from one-dimensional observations of RATAN-600 in the microwave range. \citet{Grebinskij2000} observed the circular polarization of bremsstrahlung at 17 GHz using the Nobeyama Radioheliograph (NoRH: \citealt{Nakajima1994}). They measured a longitudinal magnetic field strength of about 60--150 G in the chromosphere and corona above the active regions. Similar measurements were conducted by \citet{Iwai2013}. In addition, the above studies suggested that the circular polarization at 17 GHz contains both chromospheric and coronal components. To separate these components, additional information or assumptions are required. \citet{Iwai2014} derived the coronal component by observing the coronal loops outside of the solar limb, and measured the longitudinal magnetic field in the coronal loops as 84 G. However, this method cannot measure the coronal magnetic field in active regions on the solar disk.

The accuracy of these methods was evaluated in comparisons with observational data. However, a proper accuracy evaluation requires comparison with potential field calculations, which is difficult near the solar limb because the photospheric magnetic field cannot be accurately measured in this region. Therefore, we must derive the coronal magnetic field on the solar disk, which requires separating the coronal component from the combined chromospheric and coronal components in the thermal bremsstrahlung. The intensity of bremsstrahlung is a function of the temperature and the emission measure (hereafter, the emission measure refers to the column emission measure) of the plasma \citep{Dulk1985}. In other words, we can derive the coronal component of thermal bremsstrahlung from the coronal temperature and emission measure collected by other instruments.

The Atmospheric Imaging Assembly (AIA: \citealt{Lemen2012}) instrument onboard the {\it Solar Dynamics Observatory} ({\it SDO}) has seven EUV channels operating in six wavelengths (131, 171, 193, 211, 335, 94 {\AA}) centered on strong iron lines. The regularization code developed by \citet{Hannah2012} produces a differential emission measure (DEM) from {\it SDO}/AIA data, enabling us to derive the coronal emission measure and temperature. In this way, the temperature range was determined as $5.7<\log{T}<7.0$.

The present study estimates the accuracy of the magnetic field derived from radio observations by comparison with the potential field results and the DEM measurements using EUV observations. Section 2 describes the methods of magnetic field measurement, and Section 3 presents the observational instruments and datasets. The results of the coronal magnetic field measurements are presented in Section 4. In Section 5, we discuss the reliability of the coronal magnetic fields and compare the results with magnetic field calculations and DEM measurements. The paper is summarized in Section 6.

\section{METHODS OF MAGNETIC FIELD MEASUREMENTS}
The optical thickness of the extraordinary and ordinary modes of thermal bremsstrahlung in anisotropic plasma is computed by \citep{Gary2004}
\begin{equation}
\label{eq:tau1}
\tau_{x,o}=\int \kappa_{x,o} dl=\frac{\xi \int n_{e}\sum_{i}^{} Z_{i}^{2}n_{i}dl}{T_{e}^{3/2}(\nu \mp \nu_{B}\cos{\theta_{i}})^{2}},
\end{equation}
where $\kappa_{x,o}$ is the absorption coefficient of extraordinary and ordinary modes, $n_{e}$ is the electron density, and $Z_{i}$ and $n_{i}$ are the ion charge and number density, respectively. $T_{e}$ is the electron temperature, $\nu$ and $\nu_{B}$ denote the observational frequencies and the electron gyro-frequency, respectively, and $\theta_{i}$ is the angle between the magnetic field and the line of sight. The function $\xi$ only slightly depends on the plasma parameters, and is expressed as
\begin{equation}
\xi=9.78\times10^{-3}g_{ff},
\end{equation}
where $g_{ff}$ is the gaunt factor. Here, we adopt the gaunt factor of \citet{Gary2004}, given by 
\begin{equation}
  g_{ff} = \left\{
  \begin{array}{ll}
    18.2+\ln{T_{e}^{3/2}-\ln{\nu}} & (T_{e}>2.0\times 10^{5}\:{\rm K}) \\
    24.5+\ln{T_{e}-\ln{\nu}} & (T_{e}<2.0\times 10^{5}\:{\rm K}).
  \end{array} \right.\nonumber
\end{equation}
Assuming that the coronal plasma is fully ionized and applying the heavy ion correction \citep{Chambe1971,Gary2004}, we can use the approximation $\sum_{i} Z_{i}^{2}n_{i}\approx 1.2n_{e}$ in Eq. (\ref{eq:tau1}). In addition, using the emission measure $EM=\int n_{e}^{2}dl$, Eq. (\ref{eq:tau1}) reduces to
\begin{equation}
\label{eq:tau2}
\tau_{x,o}=\frac{1.2 \: \xi EM}{T_{e}^{3/2}(\nu \mp \nu_{B}\cos{\theta_{i}})^{2}}.
\end{equation}

The brightness temperature is determined by 
\begin{equation}
\label{eq:emis}
T_{b}^{x,o}=T_{e}[1-\exp({-\tau_{x,o}})],
\end{equation}
where $T_{b}^{x}$ and $T_{b}^{o}$ denote the brightness temperature of extraordinary and ordinary modes, respectively. Subsequently, the intensity ($I$) and polarization ($V$) components are given by 
\begin{eqnarray}
\label{eq:IV}
I&=&(T_{b}^{x}+T_{b}^{o})/2\nonumber\\
V&=&(T_{b}^{x}-T_{b}^{o})/2.
\end{eqnarray}
\citet{Bogod1980} derived the following relationship between the observed polarization degree ($V/I$) and the longitudinal magnetic field of the radiation source $B_{l}$:
\begin{equation}
\label{eq:blos}
B_{l}[{\rm G}]=\frac{10700}{n\lambda}\frac{V}{I}.
\end{equation}
In this equation, $\lambda$ is the observation wavelength and $n$ is the spectral index ($n\approx2$ in the corona), expressed as 
\begin{equation}
\label{eq:sindex}
n=\frac{\partial \log{T_{b}}}{\partial \log{\lambda}}\equiv\frac{\partial \log{I}}{\partial \log{\lambda}}.
\end{equation}
The spectral index $n$ can derived from the observed brightness temperatures at two close frequencies (here 17 GHz and 34 GHz). The spectral index $n$ is then given by
\begin{equation}
\label{eq:sindex2}
n=\frac{\log{I_{34}}-\log{I_{17}}}{\log{0.5}}, 
\end{equation}
where the subscripts $17$ and $34$ denote measurements at 17 GHz and 34 GHz, respectively.

In the microwave range, the solar atmosphere comprises two components, the optically thick chromosphere and the optically thin corona \citep{Grebinskij2000,Iwai2013}. Thus, the observed intensity and polarization components are decomposed as 
\begin{eqnarray}
\label{eq:2atmo}
I_{{\rm obs}}&=&I_{{\rm chr}}+I_{{\rm cor}}\nonumber\\
V_{{\rm obs}}&=&V_{{\rm chr}}+V_{{\rm cor}},
\end{eqnarray}
where the subscripts $obs$ , $chr$, and $cor$ indicate the observed, chromospheric, and coronal components, respectively. Having obtained the $I_{{\rm cor}}$ and $V_{{\rm cor}}$, the coronal longitudinal magnetic field can be obtained from Eq. (\ref{eq:blos}). In this study, we select the region that the chromospheric polarization component can be ignored ($V_{{\rm obs}}=V_{{\rm chr}}+V_{{\rm cor}}\approx V_{{\rm cor}}$). Hence, we can derive the coronal polarization component from the observation. The details of this assumption are discussed in Section \ref{sec:cmag}.

To determine $I_{{\rm cor}}$, we must measure the coronal temperature and emission measure. As mentioned above, \citet{Hannah2012} developed a regularization code that computes the DEM from {\it SDO}/AIA data at six EUV wavelengths (131, 171, 193, 211, 335, 94 {\AA} ). The EUV flux $F_{\lambda}$ at a given AIA wavelength $\lambda$ is calculated as 
\begin{equation}
\label{eq:AIA}
F_{\lambda}=\int\frac{dEM(T)}{dT}R_{\lambda}(T)dT,
\end{equation}
where $dEM(T)/dT$ is the DEM and $R_{\lambda}(T)$ is the filter response function. Here, the DEM is estimated by the method of least squares: 
\begin{equation}
\label{eq:chisq}
\chi^{2}=\frac{(F_{{\rm obs}}-F_{{\rm model}})^{2}}{\sigma_{F}}=\rm{min},
\end{equation}
where $\chi^{2}$ is the residual sum of squares, $F_{{\rm model}}$ and $F_{{\rm obs}}$ are the DEM-derived and observed fluxes, respectively, and $\sigma_{F}$ is the error in the observed flux. Hence, we can obtain the DEM in the temperature range of the AIA observations $(5.7<\log{T}<7.0)$.

We divided corona along the line of sight into 130 layers separated by $\log{T}=0.01$ intervals in the temperature range of the AIA observations. Therefore, the optical thickness of the $i$ th layer is
\begin{equation}
\Delta \tau_{i}=\frac{1.2\:\xi_{i}\Delta EM_{i}}{T_{e,i}^{3/2}\nu^{2}}, 
\end{equation}
where $\Delta EM_{i}$ is the integral of the DEM over the temperature range of the $i$ th layer. In the solar corona, the thermal bremsstrahlung at the microwave range is usually optically thin. Therefore, the observed brightness temperature emitted from the $i$ th layer ($\Delta T_{b,i}$) can be approximated as follows,
\begin{equation}
\label{eq:deltatb}
\Delta T_{b,i}\approx T_{e,i}[1-\exp (-\Delta \tau_{i})] \;\;\;\;\;\;\; (\tau_{i}\ll 1).
\end{equation}
The total brightness temperature is the sum of the brightness temperatures of all layers
\begin{equation}
T_{b}=\sum_{j=0}^{i-1}{\Delta T_{b,j}}.
\end{equation}

Accordingly, from the coronal temperature and AIA-based DEM calculated by the above equations, we can derive the coronal bremsstrahlung intensity $(I_{{\rm cor}}=T_{b,{\rm cor}})$ in the microwave range.

The $V_{{\rm cor}}$ is obtained by selecting the region that the chromospheric polarization component can be ignored, whereas $I_{{\rm cor}}$ is determined from the AIA observations. Hence, we can derive the coronal longitudinal magnetic field from Eq. (\ref{eq:blos}).

\section{OBSERVATIONS}
We used NoRH microwave images, EUV images and emission measure obtained by AIA, and photospheric longitudinal magnetograms collected by Helioseismic and Magnetic Imager (HMI: \citealt{Scherrer2012}). All data were acquired in the NOAA active region 11150 (S20 E02) around 3:00 UT on February 3, 2011. The photospheric longitudinal magnetic field in AR 11150 is maximized at around 1500 G. Gyro-resonance is excited at the second or third harmonics of the local gyro-frequency \citep{Shibasaki1994}. As the third harmonic of 17 GHz corresponds to a longitudinal magnetic field of about 2000 G, no gyro-resonance emission occurs at the AR 11150. The AR 11150 possesses a simple bipolar magnetic structure in the east-west direction. Soft X-ray flux was observed by the {\it Geostationary Operational Environmental Satellite} ({\it GOES}) and remained in the B-class between 20:00 UT on February 2, 2011 and 3:00 UT on February 3, 2011. Thus, no non-thermal activity such as flares occurred throughout the observation period, and we can observe only the thermal bremsstrahlung.
\subsection{Instruments}
The intensity ($I$) and polarization ($V$) images of the AR 11150 were obtained from the NoRH data. NoRH is a radio interferometer that synthesizes the radio images acquired by 84 antennas, each of 80 cm diameter, covering the full solar disk \citep{Nakajima1994}. The approximate spatial resolution is 10$\arcsec$ and 5$\arcsec$ at 17 GHz and 34 GHz, respectively, and the temporal resolution in normal mode is 1 s. At 17 GHz, NoRH can observe both the intensity ($I$) and polarization ($V$), but only the intensity ($I$) is observable at 34 GHz. EUV images and the emission measure were obtained from {\it SDO}/AIA data. The AIA simultaneously provides multiple full-disk images of the corona and transition region with spatial and temporal resolutions around 1.6$\arcsec$ and 12 s, respectively. The instrument is equipped with seven EUV filters. Six of these filters operate at wavelengths (131, 171, 193, 211, 335, and 94 {\AA}) reflecting coronal temperatures $(5.7<\log{T}<7.0)$ . From the {\it SDO}/HMI data, we obtained the photospheric longitudinal magnetic field and the vector magnetic field for the extrapolation. HMI provides photospheric magnetograms with a spatial and temporal resolution of $<1.5\arcsec$ and 45 s, respectively. For the longitudinal magnetograms, the precision is about 10 G.

\subsection{17 GHz intensity and polarization and magnetic structure}
\label{sec:17ip}
\begin{figure}
\plotone{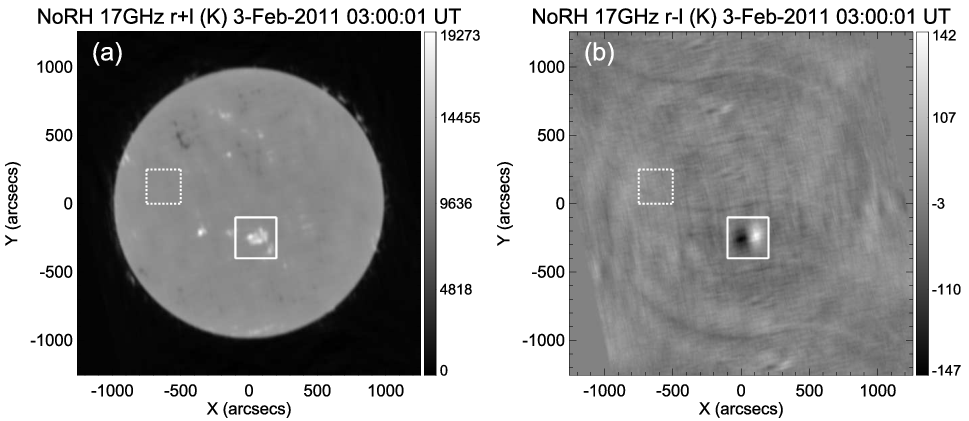}
\caption{(a) Radio intensity and (b) polarization at 17 GHz observed with NoRH around 03:00:01 UT on Feb 3, 2011. The solid and dashed rectangles enclose the NOAA active region 11150 and the quiet region, respectively.\label{fig:1}}
\end{figure}

Figure~\ref{fig:1} shows full disk images of the radio intensity and polarization observed by NoRH at 17GHz. The solid and dashed lines enclose the AR 11150 and the quiet region, respectively. Panel (a) of this figure reveals relatively strong emission in the AR 11150 region, while Panel (b) indicates a relatively strong and broadened polarized source in the same region.

To improve the signal-to-noise (S/N) ratio, we integrated 1200 images with 1 s cadence over 20 minutes. After this procedure, the standard deviation of the intensity $(\sigma_{I})$ at 17 GHz in the quiet region reduced from 120 K to 105 K, and the standard deviation of the polarization $(\sigma_{V})$ reduced from 40 K to 8 K. The minimum detectable level of the polarization degree (defined at the $5\sigma$ level) was 0.4 \%, comparable to result of \citet{Iwai2013}.

Figure~\ref{fig:2}(a) shows the photospheric longitudinal magnetic field in the AR 11150 region observed by HMI. The red and blue contours show the positive and negative components of the radio circular polarization, respectively. The locations and polarities of the photospheric longitudinal magnetic field and the radio circular polarization are strongly correlated. The positive and negative components of the polarization degree of AR 11150 were maximized at 0.9 \% and -1.1 \%, respectively.

\subsection{EUV loop structure and emission measure}
\begin{figure}
\plotone{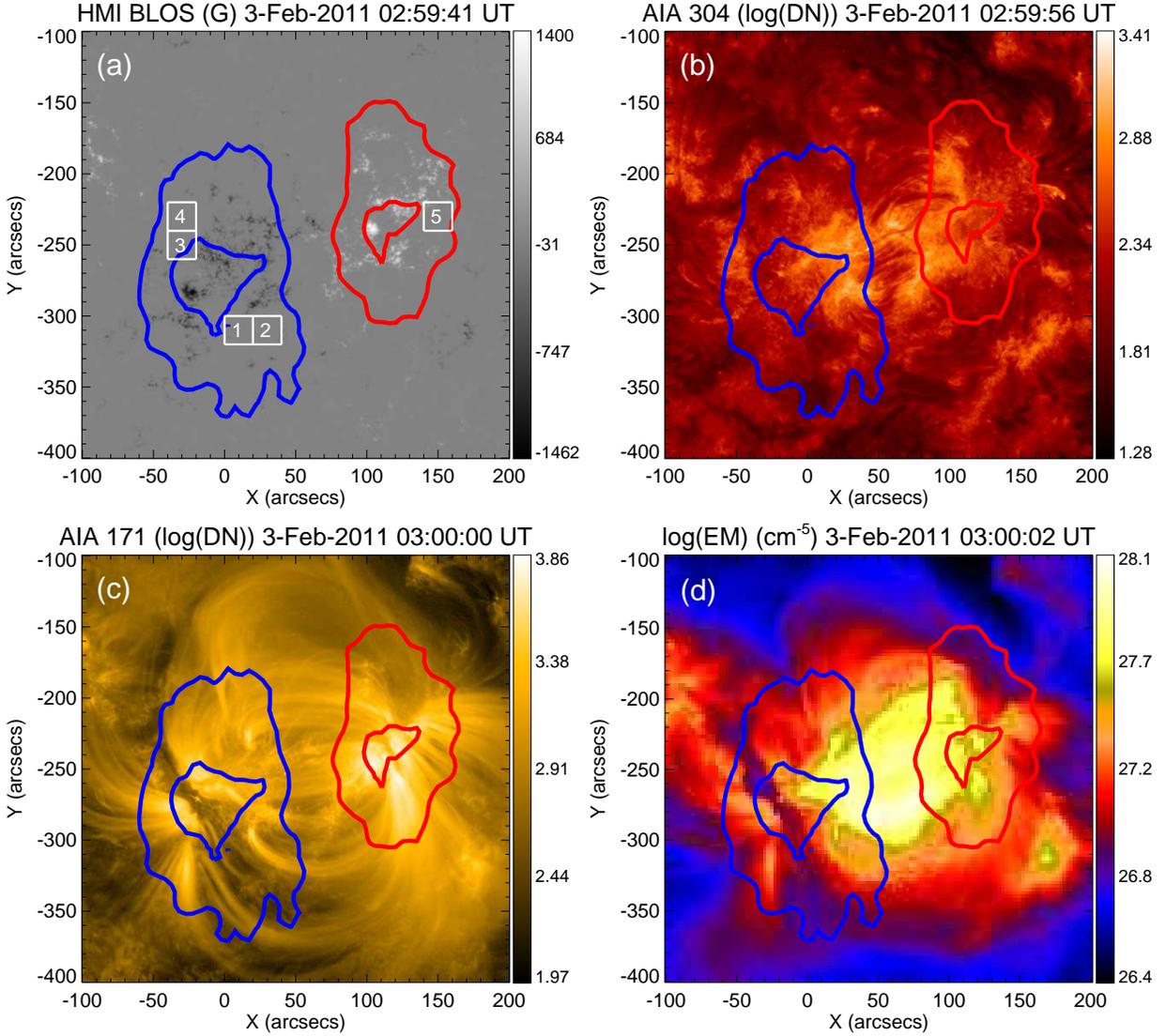}
\caption{(a) Photospheric longitudinal magnetic field observed with {\it SDO}/HMI at 03:00:35 UT on Feb 3, 2011. The coronal magnetic field was measured in the regions enclosed by the white rectangles (the measurement parameters are listed in Table~\ref{tab:1}). (b) EUV image at 304 {\AA} observed with AIA. (c) EUV image at 171 {\AA} observed with AIA. (d) Total column emission measure derived from AIA observations. The red contours show the degree of positive circular polarization (0.4 \% and 0.8 \%) at 17 GHz. The blue contours show the degree of negative circular polarization (-0.4 \% and -0.8 \%) at 17 GHz.\label{fig:2}}
\end{figure}

Figure~\ref{fig:2}(b) presents the EUV image observed by AIA at 304 {\AA}. The 304 {\AA} line is emitted from the upper chromosphere and the transition region at approximately $5\times 10^{4}$ K. There is a relatively strong emission at the AR 11150 center. However, the radio polarization and the EUV emission at 304 {\AA} do not spatially coincide. Figure~\ref{fig:2}(c) shows the EUV image observed by AIA at 171 \AA. The 171 {\AA} line is emitted from the corona at about $10^{6}$ K. The polarization source is distributed near the foot of the coronal loops, as expected from knowledge that the polarization degree of the thermal bremsstrahlung depends on the longitudinal magnetic field. In fact, at the coronal loops tilted from the line of sight (such as the loop top), the polarization degree never exceeds the minimum detectable level of 0.4 \%.

The DEM was calculated from the AIA data using Eqs. (\ref{eq:AIA}) and (\ref{eq:chisq}). Figure~\ref{fig:2}(d) plots the total EM ($=\int{{\rm DEM}}dT$), obtained by integrating the DEM over the temperature range of AIA $(5.7<\log{T}<7.0)$. To improve the S/N ratio, the AIA data are integrated with an area of $4\times4$ pixels. Large emission measure (${\rm EM}\geq 10^{27}{\rm cm}^{-5}$) is observed at the AR 11150 center.

\subsection{Potential field model}
We used a potential field source surface model \citep{Schatten1969,Altschuler1969,Shiota2008,Shiota2012} to extrapolate coronal magnetic fields from the photospheric magnetic field observations. These model requires a radial magnetic field on the whole surface of the Sun. We used synoptic maps of the radial magnetic field in the Carrington rotation 2106.5 produced from the HMI observation. The synoptic data around the AR 11150 was replaced by the radial components of a vectormagnetogram of the HMI/SHARP series \citep{Bobra2014} at 3:00 UT on February 2011. The synthetic synoptic map was expanded with spherical harmonics of degree 1800. The three-dimentional coronal magnetic field was reconstructed assuming that all the field lines open out at 2.5 solar radii. The spatial resolution of this model is about 2400 km. This high spatial resolution enables the potential field model to reproduce fine magnetic structures near the surface such as canopy structure rooted on the network fields \citep{Ito2010,Shiota2012}. 

\section{RESULTS}
\label{sec:cmag}
The coronal longitudinal magnetic fields were derived from regions satisfying the following conditions. First, we selected regions exhibiting weak photospheric longitudinal magnetic fields $(\approx 10\:{\rm G})$. Therefore, we can neglect the chromospheric polarization component at 17 GHz. Second, we selected the region with many coronal loops. As shown in Figure~\ref{fig:2}(c), such a region is found around the center of AR 11150, and indicates the presence of a coronal magnetic field. Third, we selected regions in which the polarization degree exceeded the minimum detectable level at 17 GHz ($=0.4 \%$).

The five regions satisfying the above conditions are numbered and enclosed in white rectangles in Figure~\ref{fig:2}(a) (hereafter, these regions are called the numbered regions). The sizes of these regions (20$\arcsec$) exceed the beam size at 17 GHz $(\approx 10-15\arcsec)$.

Table~\ref{tab:1} tabulates the measurement parameters of the numbered regions depicted in Figure~\ref{fig:2}(a). The observed intensities at 17 GHz ($I_{{\rm obs,17}}$) range from 10400 K to 11200 K, relatively higher than the typical intensity of the quiet region $(\approx 10000\:{\rm K})$. The coronal intensities at 17 GHz ($I_{{\rm cor,17}}$) in the numbered regions, derived by AIA observations, are 700--1400 K. Thus, the corresponding chromospheric intensities $(I_{{\rm chr,17}}\approx I_{{\rm obs,17}}-I_{{\rm cor,17}})$ at 17 GHz are 9700--10400 K.

To derive the spectral index $n$, we also observed the intensity at 34 GHz ($I_{{\rm obs,34}}$). In addition, we derived the chromospheric and coronal intensities at 34 GHz ($I_{{\rm chr,34}}$ and $I_{{\rm cor,34}}$ respectively) by the method used to derive their 17 GHz counterparts. From these results, we derived the spectral index at the chromosphere and corona from Eq. (\ref{eq:sindex2}). The values are listed in Table~\ref{tab:1}. The coronal spectral indexes $n_{{\rm cor}}$ in the numbered regions are consistent with the typical coronal value (approximately 2). In contrast, the chromospheric spectral indexes $n_{{\rm chr}}$ are an order of magnitude lower than the coronal spectral indexes. Consequently, to maintain the same polarization degree, the chromospheric longitudinal magnetic field strength must be an order of magnitude greater than the coronal field strength. Therefore, it is suitable to use the approximation $V_{{\rm obs}} \approx V_{{\rm cor}}$ in the numbered regions.

As demonstrated in Table~\ref{tab:1}, the photospheric magnetic fields $B_{l,{\rm pho}}$ in the numbered regions are very weak ($\approx$ 10 G). The magnetic structures of the photosphere and chromosphere are assumed similar at the spatial resolution scale of 17 GHz ($\approx 10\arcsec$). Thus, the polarization components of the chromosphere $V_{{\rm chr}}$ in the numbered regions are negligible, and the observed polarization $V_{{\rm obs}}$ is comparable with $V_{{\rm cor}}$. The derived parameters $I_{{\rm cor,17}}$ and $n_{{\rm cor}}$ are also listed in Table~\ref{tab:1}. Inserting $I_{{\rm cor,17}}$, $V_{{\rm cor}}$, $n_{{\rm cor}}$ into Eq. (\ref{eq:blos}), the coronal longitudinal magnetic field $B_{l,{\rm cor}}$ was computed as around 150--270 G in the numbered regions. 

We now estimate the measurement error in the coronal longitudinal magnetic field derived from radio and EUV observations. To this end, we investigate how the errors combine in Eq. (\ref{eq:blos}). The error in the coronal longitudinal magnetic field $\sigma_{B_{l,{\rm cor}}}$ becomes
\begin{equation}
\sigma_{B_{l,{\rm cor}}}=\frac{10700}{n\lambda}\sqrt{\frac{\sigma_{V_{\rm cor}}^{2}}{I_{\rm cor,17}^{2}}+\frac{V_{\rm cor}^2\sigma_{I_{\rm cor,17}}^{2}}{I_{\rm cor,17}^{4}}},
\end{equation}
where $\sigma_{V_{{\rm cor}}}$ and $\sigma_{I_{{\rm cor,17}}}$ are the errors in $V_{{\rm cor}}$ and $I_{{\rm cor,17}}$, respectively. Since we assumed $V_{{\rm obs}} \approx V_{{\rm cor}}$ in the numbered regions, $\sigma_{V_{{\rm cor}}}$ denotes the error in the observed polarization component $\sigma_{V_{{\rm obs}}}$. In Section \ref{sec:17ip}, this error was determined as $\sigma_{V_{{\rm obs}}}=$ 8 K. 

To estimate $I_{{\rm cor,17}}$, we used the DEM derived from the AIA observations. Since $I_{{\rm cor,17}}$ is proportional to the emission measure, we must derive the error in the emission measure. 

\begin{figure}
\plotone{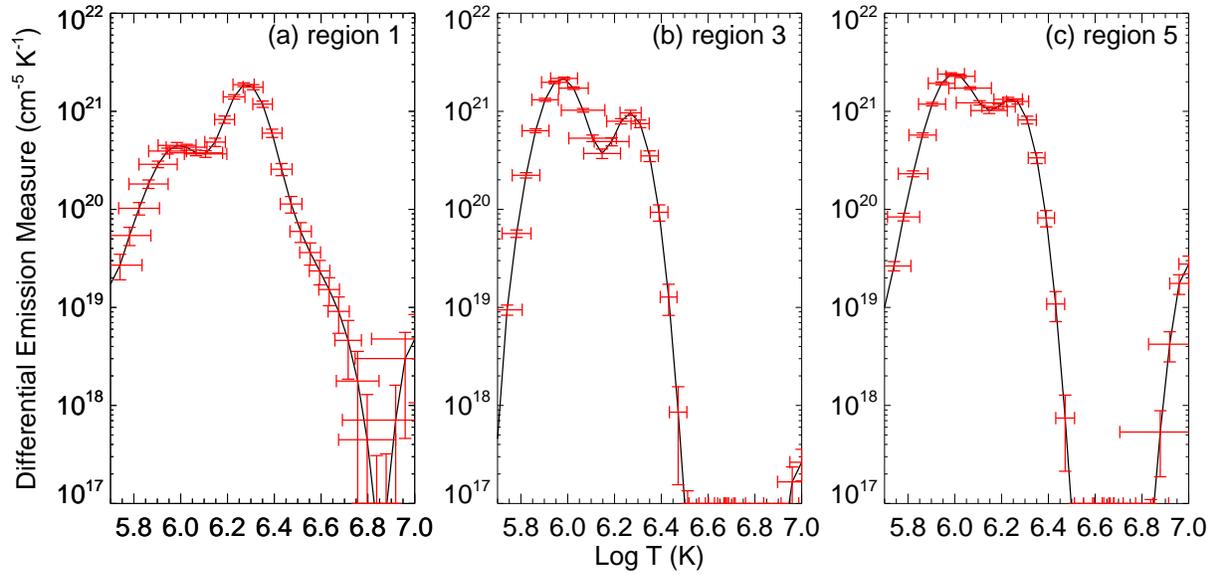}
\caption{Comparison of the differential emission measure derived from AIA in different regions; (a) Region 1, (b) Region 3, and (c) Region 5. The error bars are calculated by standard Monte Carlo approach used in \citet{Hannah2012}. \label{fig:3.5}}
\end{figure}

Figure~\ref{fig:3.5} shows the DEM derived from AIA in Regions 1, 3, and 5. The error of the DEM is calculated by standard Monte Carlo approach used in \citet{Hannah2012}. In the low-temperature range ($\log{T} < 6.5$), the error of the DEM is up to about 10 \%. In contrast, the error of the DEM in the high-temperature range ($\log{T} > 6.5$) is greater than that of the low-temperature range. However, the DEM in the high-temperature range are more than two orders of magnitude lower than the low-temperature range. Therefore, the effect of the error of the DEM in the high-temperature is negligible.

We derived the DEM by the least squares method, adopting a single-degree-of-freedom (SDF) with a significance level of ($\chi^{2}=3.841$). The chi-squared $\chi^{2}$ of the numbered regions in this study was close to 2, implying that the DEM of the numbered regions is significant.

For these reasons, we estimated the error in the emission measure as about 10 \%. In this case, the error in the coronal longitudinal magnetic field was estimated as 20--40 G, which is about 15 \% of the measured value.

\section{DISCUSSION}
\label{sec:5}
\subsection{Comparison with the potential field model}
One of the purpose of this study was to validate the accuracy of magnetic fields derived from radio observations by comparing them with potential field calculations. Since the potential field is difficult to calculate near the solar limb, we derived the coronal magnetic field on the solar disk. This section compares the coronal longitudinal magnetic fields with the potential field \citep{Shiota2008,Shiota2012}, and assesses their accuracy.

\begin{figure}
\plotone{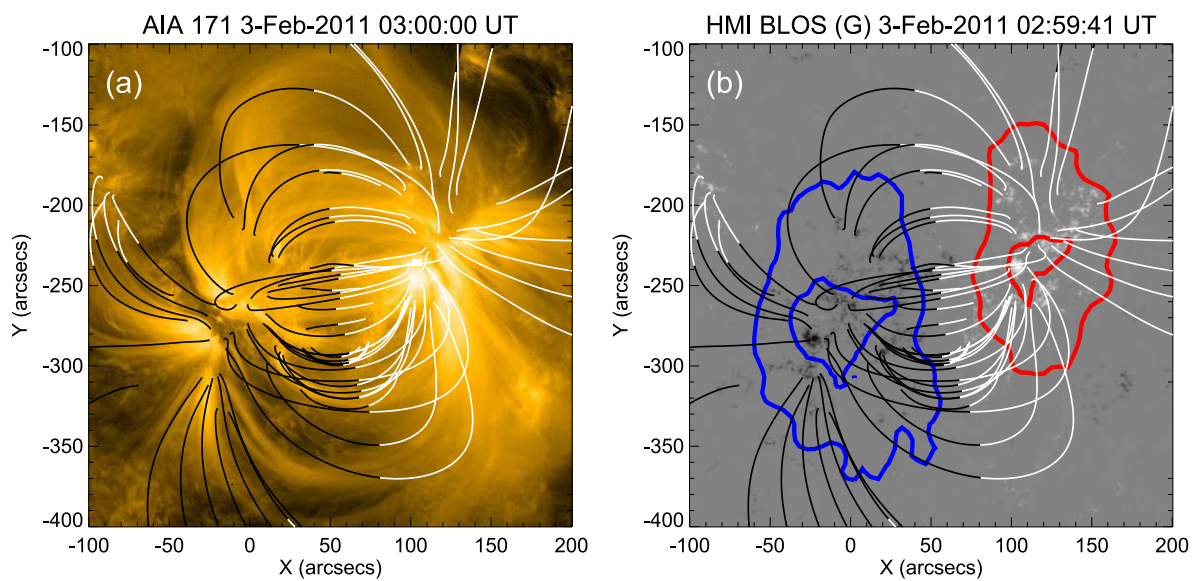}
\caption{(a) EUV image at 171 {\AA} observed with AIA. (b) Photospheric longitudinal magnetic field observed with {\it SDO}/HMI at 03:00:35 UT on Feb 3, 2011. The red contours show the degree of positive circular polarization (0.4 \% and 0.8\%) at 17 GHz. The blue contours show the degree of negative circular polarization (-0.4 \% and -0.8 \%) at 17 GHz. The white and black lines show outward and inward (radially positive and negative) parts of magnetic field lines of the potential field, respectively. \label{fig:3}}
\end{figure}

Figure~\ref{fig:3} shows the magnetic field lines of the potential field. The coronal loops of EUV structurally agree with the magnetic field lines of the potential field (Figure~\ref{fig:3}(a)). Therefore, the magnetic fields of the AR 11150 can be considered to be a potential like. In Figure~\ref{fig:3}(b), the polarization source is distributed near the foot of the magnetic field lines, consistent with the dependence of the polarization degree of the thermal bremsstrahlung on the longitudinal magnetic field. 

\begin{figure}
\epsscale{0.7}
\plotone{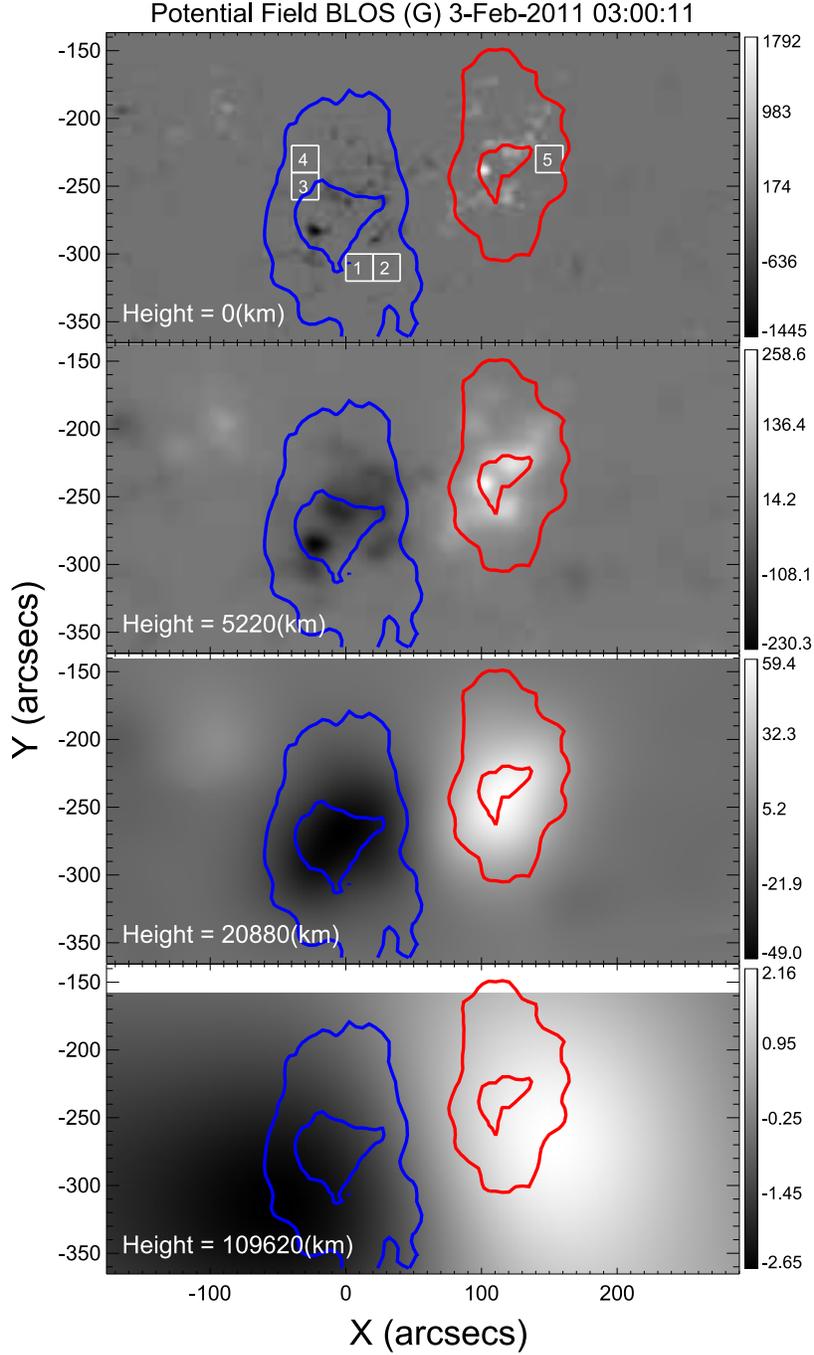}
\caption{Longitudinal components of the potential field in the AR 11150 at heights of 0 km (top), 5220 km (upper center), 20880 km (lower center) and 109620 km (bottom) above the photosphere. The red contours show the degree of positive circular polarization (0.4 \% and 0.8 \%) at 17 GHz. The blue contours show the degree of negative circular polarization (-0.4 \% and -0.8 \%) at 17 GHz. The regions in which the coronal magnetic field were measured are enclosed in the white rectangles, and the parameters are described in Table~\ref{tab:1}. \label{fig:4}}
\end{figure}

Figure~\ref{fig:4} shows the longitudinal component of the potential field in the AR 11150. The height refers to the altitude from the photosphere surface. The magnetic field at zero height is the photospheric longitudinal magnetic field measured by HMI. This field does not fill the radio polarized region at 5220 km height, which corresponds to the bottom of the corona in the atmospheric model of \citet{Selhorst2005}. On the other hand, at 20880 km height, the longitudinal magnetic field appears to be distributed throughout the contour of the polarization degree. However, the maximum magnetic field strength at this height ($\approx 60\:{\rm G}$) is weaker than the coronal longitudinal magnetic field strength derived in this study. In addition, although the longitudinal magnetic field adequately fills the radio polarized region at 109620 km, the magnetic field strength at this height is very weak ($<2\:{\rm G}$).

\begin{figure}
\epsscale{1.0}
\plotone{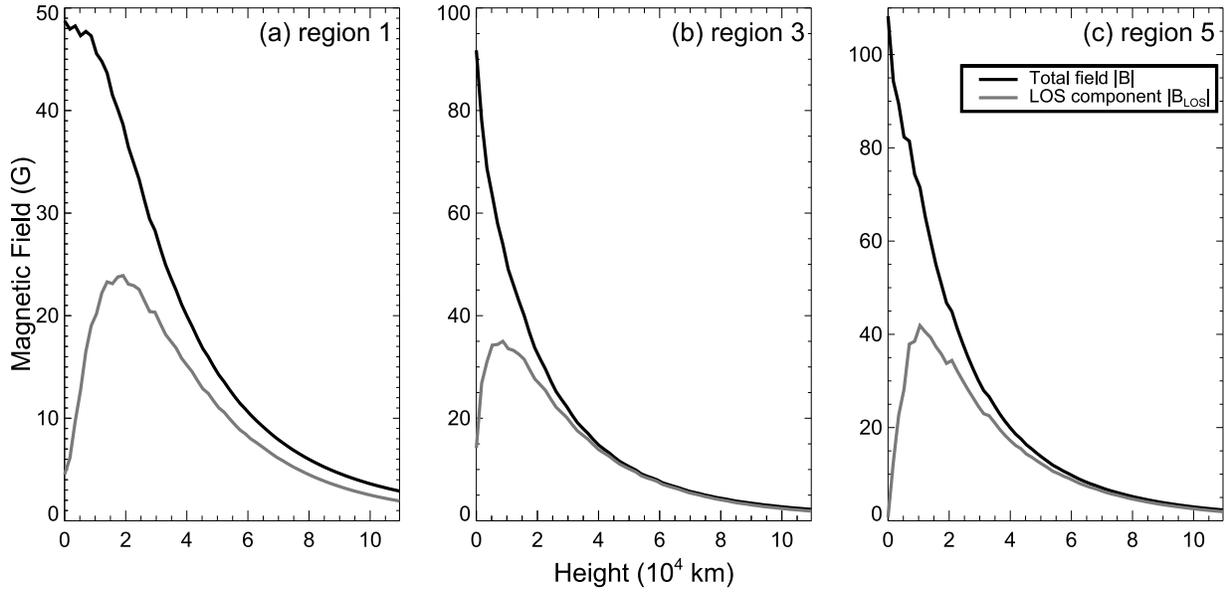}
\caption{Comparison of the height variation of the potential field strength in different regions; (a) Region 1, (b) Region 3, and (c) Region 5. Black and gray lines show the absolute values of the magnetic field strength $|B|=\sqrt{B_{x}^{2}+B_{y}^{2}+B_{z}^{2}}$ and the longitudinal magnetic field strength $|B_{l}|$, respectively.\label{fig:5}}
\end{figure}

The height variations of the potential field strength in Regions 1, 3, and 5 are shown in Figure~\ref{fig:5}. The longitudinal magnetic fields monotonically increase with height, up to approximately 10000--20000 km. At the upper heights, a clear peak appears in the longitudinal magnetic field strength, which corresponds to the corona. Accordingly, the magnetic field at each coronal region is considered to be dominated by the magnetic field lines component that broadens from the AR 11150 center. At 10000--20000 km, the absolute value of the longitudinal component $|B_{l}|$ of the potential field peaks at 24, 30, and 45 G in Regions 1, 3 and 5, respectively. On the other hand, the measured values of the coronal longitudinal magnetic fields $|B_{l,{\rm cor}}|$ are 272, 276, and 166 G, respectively, a whole order of magnitude greater than the potential fields. In Section \ref{sec:cmag}, the measurement error was determined as approximately 15\%. Even accounting for this error, the derived magnetic field exceeds the potential field.

\subsection{Upper limit of the derived coronal magnetic field}
The measured magnetic field differs from the potential field chiefly because the coronal emission measure is underestimated. The temperature range of the AIA observations, from which we derived the coronal intensity $I_{{\rm cor,17}}$, was $5.7<\log{T}<7.0$. This range excludes the low-temperature coronal plasma ($\log{T}<5.7$). \citet{Iwai2014} derived the emission measure of coronal loops outside the solar limb using both NoRH and AIA data. They reported a greater emission measure derived from NoRH than from AIA. According to the 17 GHz NoRH data, all of the plasma in the coronal loops outside the solar limb is optically thin, indicating a low-temperature coronal plasma in these regions ($\log{T}<5.7$). As \citet{Iwai2014} derived the emission measure by a method of \citet{Aschwanden2013}, which is similar to ours, we infer that the coronal emission measure is similarly underestimated in the present study. We therefore assign our derived coronal intensity $I_{{\rm cor,17}}$ as a lower limit. For these reasons, we also consider the coronal longitudinal magnetic field $B_{l,{\rm cor}}$ calculated by Eq. (\ref{eq:blos}) as an upper limit.

\subsection{The effect of the low-temperature plasma}
\label{sec:cmagatomo}
\begin{figure}
\plotone{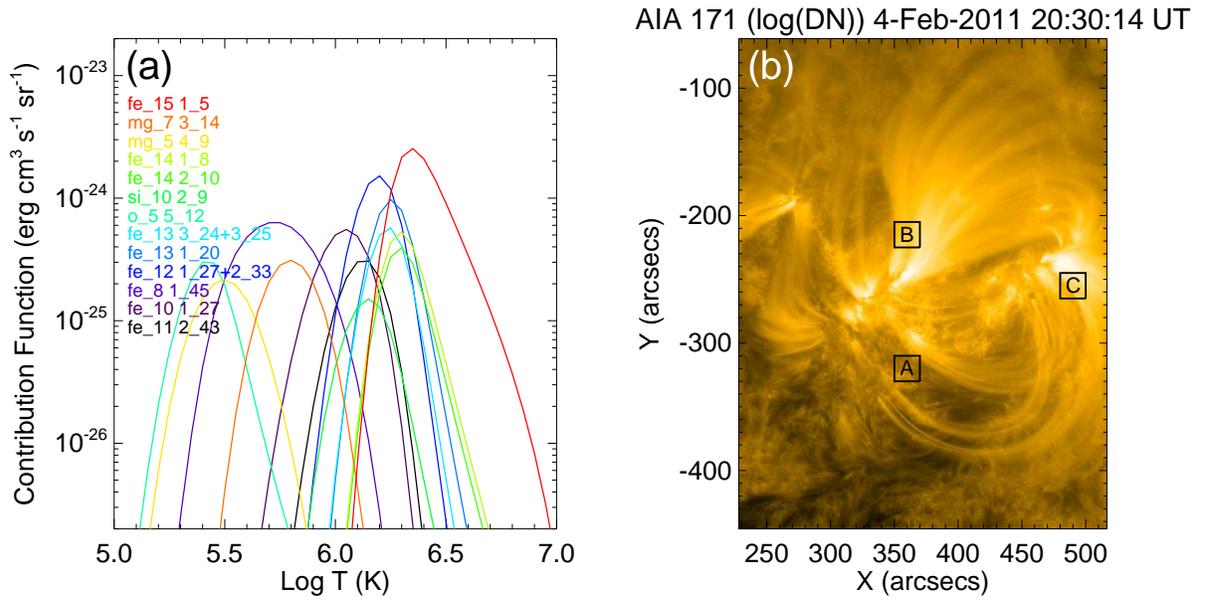}
\caption{(a) Contribution functions for several EUV lines observed with Hinode/EIS. (b) EUV image at 171{\AA} observed with AIA at 20:30:14 UT on Feb 4, 2011. The regions A, B and C enclosed by black rectangles correspond to the regions 1, 3 and 5 in Figure~\ref{fig:2}(a), respectively. \label{fig:7}}
\end{figure}

As explained above, we treat the coronal magnetic field strength derived in this study as an upper limit. To estimate the actual coronal magnetic field strength, we should evaluate the effect of the low-temperature plasma. The EUV Imaging Spectrometer (EIS: \citealt{Culhane2007}) instrument onboard the {\it Hinode} satellite can observe some low-temperature ($\log{T}<5.7$) sensitive lines. Unfortunately, there are no EIS data of the AR 11150 on February 3. Hence, we used the EIS data of the AR 11150 obtained between 20:03--21:05 UT on February 4, 2011, and compared them with the AIA data obtained in the same period. Note that these data were from the same active region as shown in Figure~\ref{fig:2}, but obtained about 41.5 hours later. Figure~\ref{fig:7}(a) shows the contribution functions of the lines observed with EIS in this study. The contribution functions are calculated with the CHIANTI version 7.1.3, a constant pressure assumption of $10^{-15}\:{\rm cm^{-3}\:K}$, and a set of coronal elemental abundances \citep{Feldman1992}. Two lines which sensitive to the low-temperature range, O V ($\log{T_{{\rm max}}}\sim5.4$) and Mg V ($\log{T_{{\rm max}}}\sim5.5$) are included in the EIS data in this study. Figure~\ref{fig:7}(b) shows the EUV image at 171 {\AA} observed with AIA. We defined the regions A, B and C enclosed by black rectangles that correspond to the regions 1, 3 and 5 in Figure~\ref{fig:2}(a), respectively.

\begin{figure}
\plotone{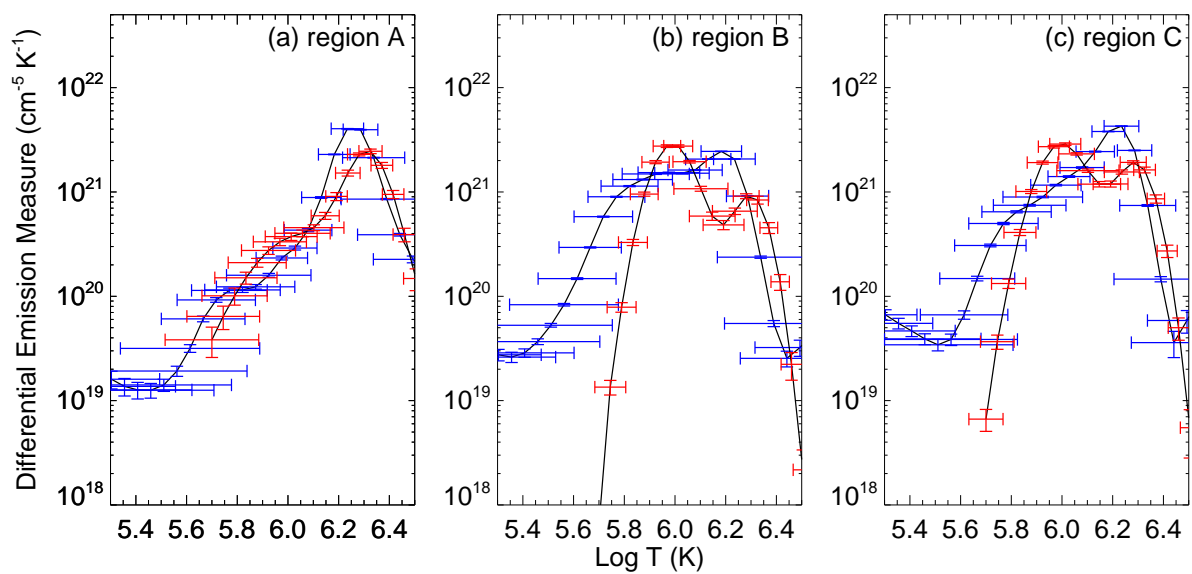}
\caption{Comparison of the differential emission measure derived from AIA (red error bar) and EIS (blue error bar) respectively, in different regions; (a) Region A, (b) Region B, and (c) Region C that correspond to the regions 1, 3 and 5 in Figure~\ref{fig:2}(a), respectively. The error bars are calculated by standard Monte Carlo approach used in \citet{Hannah2012}.\label{fig:8}}
\end{figure}

Figure~\ref{fig:8} shows the DEM derived from AIA and EIS at the regions A, B and C. Because of the limitation of the lines observed by EIS, we calculated the DEM in the temperature range of $5.3 < \log{T} < 6.5$ from the EIS data. 

\begin{figure}
\plotone{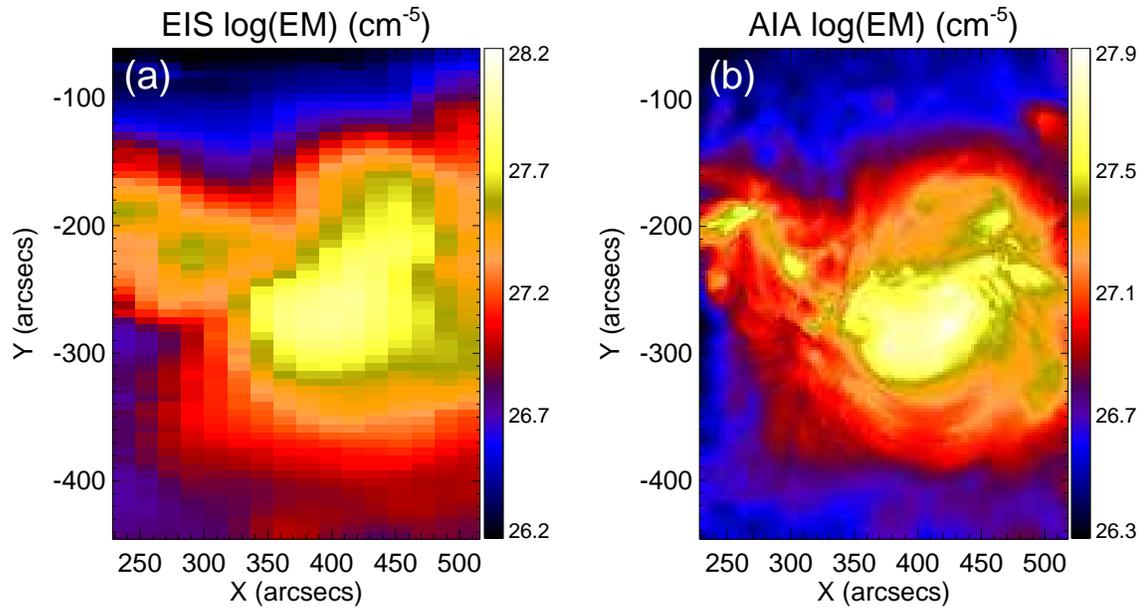}
\caption{(a, b) Total column emission measure derived from EIS and AIA observations at around 20:30 UT on Feb 4, 2011, respectively.\label{fig:9}}
\end{figure}

Figures~\ref{fig:9}(a) and (b) show the emission measure obtained by EIS and AIA respectively. To improve the S/N ratio, the EIS and AIA data are integrated with an area of $6\times6$ pixels and $4\times4$ pixels, respectively. These total EM are integrated in the temperature range of $5.3 <  \log{T} < 6.5$ for the DEM derived from EIS, and $5.7 < \log{T} < 6.5$ for the DEM of derived from AIA. Table~\ref{tab:2} shows the total EM at the regions indicated in Figure~\ref{fig:7}(b). The emission measures derived from EIS ($EM_{\rm EIS}$) are greater than that of derived from AIA ($EM_{\rm AIA}$) by about 20--35 \%. This is because the $EM_{\rm EIS}$ includes the low-temperature range ($5.3 < \log{T} < 5.7$). A discrepancy of the DEM derived from EIS and AIA at the temperature of $\log{T}\sim6.2$ in Figure~\ref{fig:8} is also affect the difference of the EM derived from EIS and AIA. Since the DEM is calculated by the least squared method, and $\chi^2$ derived from EIS and AIA at these regions are too small, the DEM of these regions are significant. Therefore, the total EM derived from EIS is certainly greater than AIA, implying the presence of low-temperature plasma of about 20--35 \% at these regions. If the EM derived from AIA is underestimated about 35 \%, the coronal magnetic field derived in the numbered regions should be 100--210 G. Since the DEM derived from EIS excludes the temperature range of  $\log{T} < 5.3$, these magnetic fields should also be an upper limit. In order to specify the effect of the low-temperature coronal plasma more accurately, the DEM analysis using lines which sensitive to the lower temperature range ($\log{T} < 5.3$) are needed for future studies.

Since the derived coronal longitudinal magnetic fields estimated from EIS observations were still the upper limit, we should estimate the actual coronal magnetic field by using some assumptions.
The millimeter-range brightness temperatures in the quiet region have been observed (e.g. \citealt{Kuseki1976}) and modeled \citep{Selhorst2005}. According to \citet{Selhorst2005}, the brightness temperature of the quiet region at 34 GHz is about 9000 K. The height of the optically thick layer at 17 GHz should be higher than that of at 34 GHz because the temperature in the chromosphere is a monotonically increasing function of height. Hence, the chromospheric intensity at 17 GHz $I_{{\rm chr,17}}$ should be greater than 9000 K. If we assume that the chromospheric intensity at 17 GHz is 9000 K, the coronal intensities $I_{{\rm cor,17}}$ of the numbered regions range from 1500 K to 2500 K. Accordingly, the coronal longitudinal magnetic field is estimated to be 90--130 G, which is still greater than the potential field strength.

\subsection{Underestimation of the potential field}
We consider several reasons for the underestimated potential field strength on the basis of the HMI observation. First, the HMI instrument has an intrinsic systematic underestimation of the magnetic field measurement because of the magnetic field sensitivity of the spectal Line Fe {\small I} 6173 {\AA} used in this instrument. The less sensitivity has been studied with cross calibration with other observations. \citet{Liu2012} compared with HMI magnetic field observations with simultaneous observations obtained with the Michelson Doppler Imager (MDI) aboard {\it Solar and Heliospheric Observatory} ({\it SOHO}) and showed that the magnetic field strength (flux density) observed by HMI is 1.44 times smaller than that of observed by MDI within the weak field range near the disk center ($B_{\rm MDI} \sim 1.44 B_{\rm HMI}$). The MDI magnetic field observations have been compared with other observations \citep{Berger2003,Wang2009}. \citet{Wang2009} investigated the magnetic field observations of MDI and Spectropolarimeter (SP) aboard the {\it Hinode} satellite. They showed that the averaged ratio of the observed magnetic field between MDI and SP is $0.71 \pm 0.09$ $(B_{\rm MDI} \sim 0.71 B_{\rm SP})$. These studies suggest that the magnetic field strength observed with HMI can be about half of the magnetic field strength observed by the SP. If we assume that the actual magnetic field is twice larger than the magnetic field observed by HMI, the calculated peak strengths of the coronal potential magnetic field are 48--90 G for Regions 1, 3, 5. These values are still less than those of discussed in Section \ref{sec:cmagatomo}.

The magnetic field between the photosphere and chromosphere is considered to be not current-free \citep{Gary2001}, because the corresponding plasma has a high beta value. Hence, the accurate coronal field should not be derived from the potential field model with a photospheric magnetic field. The potential field is just a reference magnetic field structure in the minimum state of energy and therefore the actual coronal magnetic field is expected to have different (probably higher) strength than in the potential model. In addition to the difference in strength, the configuration should be different. Because the potential field does not contain current, the field very close to the solar surface becomes more horizontal. However, the actual field structures tend to be open (forming funnel structures) as inferred from H$\alpha$ spicule observations \citep{Giovanelli1980}. Hence, the actual magnetic field can contain more radial component than the potential field.

From the reasons mentioned above, the actual magnetic field strength is expected to be larger than that of the potential field.

\section{SUMMARY}
We derived the coronal magnetic field from the polarization observation of the radio thermal bremsstrahlung. Magnetic field measurements based on the circularly polarized thermal bremsstrahlung have been investigated by several studies \citep{Grebinskij2000,Iwai2013}. However, the circularly polarization observed at 17 GHz contains both chromospheric and coronal components. Therefore, previous studies assumed the chromospheric brightness temperature such as $I_{{\rm chr}}=10000\:{\rm K}$. In contrast, we separated coronal and chromospheric components by combining the radio and EUV observations and derived the coronal longitudinal magnetic field. In addition, we evaluated the derived magnetic field by comparing it with potential magnetic field and the DEM measurements using EUV observations. Our conclusions are summarized as follows; 
\begin{enumerate}
\item The coronal longitudinal magnetic fields in the AR 11150 were estimated as 150--270 G from NoRH and AIA observations. These values were weighted by the coronal emission measure over the line-of-sight.
\item The longitudinal components of the potential field in AR 11150 were strongest (24--45 G) at 10000--20000 km above the photosphere surface. These values are an order of magnitude smaller than the coronal longitudinal magnetic field derived in this study.
\item As AIA observes over a limited temperature range $(5.7<\log{T}<7.0)$, we took the coronal radio intensity $I_{{\rm cor,17}}$ derived at 17 GHz as the lower limit. Hence, the derived coronal longitudinal magnetic field $B_{l,{\rm cor}}$ was considered as the upper limit. The measurement error in the upper limit was estimated at 15 \%.
\item We also derived the differential emission measure including the low-temperature plasma $5.3<\log{T}<5.7$ from the {\it Hinode}/EIS observation. In this case the upper limit of the coronal longitudinal magnetic fields in the AR 11150 were estimated to be 100--210 G.
\item We estimated the coronal magnetic field with some assumptions. If the chromospheric radio intensity at 17 GHz is 9000 K, the coronal longitudinal magnetic field is estimated as 90--130 G, which still exceeds the magnetic field derived from the potential field.
\item The discrepancy between the potential and the observed magnetic field strengths can be explained consistently by two reasons; (a) the underestimation of the coronal emission measure resulting from the limitation of the temperature range of AIA and EIS, (b) the underestimation of the coronal longitudinal magnetic field resulting from the potential field assumption.
\end{enumerate}
We derived the coronal longitudinal magnetic field from simultaneous NoRH and AIA observations and using the DEM derived from the EIS. However, the coronal longitudinal magnetic field strength derived in this study was an upper limit because of the limited temperature range of the AIA and EIS observations. To more accurately determine the coronal magnetic field by our method, new instruments or methods for calculating the differential emission measure of low-temperature coronal plasma are required. In addition, improving the S/N ratio of the radio interferometer would reduce the measurement error in the coronal magnetic field.

\acknowledgments
The authors are grateful to Takashi Sakurai and Shinsuke Imada for useful discussions. {\it SDO} data and images are courtesy of NASA/{\it SDO} and the AIA and the HMI science team. This work was conducted by a joint research program of the National Astronomical Observatory Japan. This work was partially supported by JSPS KAKENHI Grant Number 26800255.

%%%%%%%%%%%%%%%%%%%%%%%%%%%%%%%%%%%%%%%%%%%%%%%%%%%%%%%%%

\input{table1_revise2.tex}
\input{table2.tex}
\end{document}

%% file: table1_revise2.tex
\begin{table}
\scalebox{0.75}[0.75]
{
\begin{tabular}{c|ccccccccccc|cc}
\tableline
& $I_{{\rm obs,17}}$ & $I_{{\rm chr,17}}$ & $I_{{\rm cor,17}}$ &  $I_{{\rm obs,34}}$ & $I_{{\rm chr,34}}$ & $I_{{\rm cor,34}}$ & $V_{{\rm obs}}$ & $n_{{\rm chr}}$ & $n_{{\rm cor}}$ & $EM_{{\rm cor}}$ & $B_{l,{\rm pho}}$ & $B_{l,{\rm cor}}$ & $\sigma_{B_{l,{\rm cor}}}$\\
Region & (K) & (K) & (K) & (K) & (K) & (K) & (K) & & & $(10^{27}{\rm cm^{-5}})$ & (G) & (G) & (G)\\
\tableline
1 & 11247 & 10395 & 852 & 9819 & 9615 & 203 & -79 & 0.11 & 2.07 & 1.84 &  -1 & -272 & 40 \\
2 & 11505 & 10094 & 1411 & 9715 & 9377 & 336 & -71 & 0.10 & 2.06 & 3.05 &  -1 & -147 & 23\\
3 & 10953 & 10143 & 810 & 9421 & 9227 & 193 & -76 & 0.13 & 2.06 & 1.48 &  -9 & -276 & 41 \\
4 & 10457 & 9661 & 796 & 9072 & 8881 & 190 & -63 & 0.12 & 2.06 & 1.49 &  -1 & -232 & 39 \\
5 & 10898 & 9794 & 1104 & 9783 & 9519 & 262 & 63 & 0.03 & 2.07 & 2.07 &  16 & 166 & 28 \\
\tableline
\end{tabular}
}
\caption{Measurement parameters of the numbered regions indicated in Figure 2(a)\label{tab:1}}
\end{table}

%% file: table2.tex
\begin{table}
{
\begin{tabular}{c|cc|c}
\tableline
Region & $EM_{{\rm EIS}}\:({\rm cm^{-5}})$ & $EM_{{\rm AIA}}\:({\rm cm^{-5}})$  & $EM_{{\rm EIS}}$ / $EM_{{\rm AIA}}$\\
\tableline
A & $3.46\times10^{27}$ & $2.56\times10^{27}$ & 1.35 \\
B & $2.59\times10^{27}$ & $1.99\times10^{27}$ & 1.30 \\
C & $3.63\times10^{27}$ & $3.10\times10^{27}$ & 1.21  \\
\tableline
\end{tabular}
}
\caption{The total EM at the regions indicated in Figure 7(b)\label{tab:2}}
\end{table}